\documentstyle[twoside]{article}

\catcode`\@=11
\long\def\@makefntext#1{ %\parindent 1em
\protect\noindent \hbox to 3.2pt {\hskip-.9pt
$^{{\eightrm\@thefnmark}}$\hfil}#1\hfill} %can be used

\def\thefootnote{\fnsymbol{footnote}}
 \def\@makefnmark{\hbox to 0pt{$^{\@thefnmark}$\hss}}  %original

\def\ps@myheadings{\let\@mkboth\@gobbletwo
\def\@oddhead{\hbox{} %\sl
\rightmark\hfil\eightrm\thepage}
\def\@oddfoot{}\def\@evenhead{\eightrm\thepage\hfil %\sl
\leftmark\hbox{}}\def\@evenfoot{}
\def\sectionmark##1{}\def\subsectionmark##1{}}

\oddsidemargin=\evensidemargin
\renewcommand{\thefootnote}{\fnsymbol{footnote}}

\newcounter{sectionc}\newcounter{subsectionc}\newcounter{subsubsectionc}
\renewcommand{\section}[1] {\vspace{12pt}\addtocounter{sectionc}{1}
\setcounter{subsectionc}{0}\setcounter{subsubsectionc}{0}\noindent
	{\tenbf\thesectionc. #1}\par\vspace{5pt}}
\renewcommand{\subsection}[1] {\vspace{12pt}\addtocounter{subsectionc}{1}
	\setcounter{subsubsectionc}{0}\noindent
	{\bf\thesectionc.\thesubsectionc. {\kern1pt \bfit #1}}\par\vspace{5pt}}
\renewcommand{\subsubsection}[1]
{\vspace{12pt}\addtocounter{subsubsectionc}{1}
	\noindent{\tenrm\thesectionc.\thesubsectionc.\thesubsubsectionc.
	{\kern1pt \tenit #1}}\par\vspace{5pt}}
\newcommand{\nonumsection}[1] {\vspace{12pt}\noindent{\tenbf #1}
	\par\vspace{5pt}}

\newcounter{appendixc}
\newcounter{subappendixc}[appendixc]
\newcounter{subsubappendixc}[subappendixc]
\renewcommand{\thesubappendixc}{\Alph{appendixc}.\arabic{subappendixc}}
\renewcommand{\thesubsubappendixc}
	{\Alph{appendixc}.\arabic{subappendixc}.\arabic{subsubappendixc}}

\renewcommand{\appendix}[1] {\vspace{12pt}
        \refstepcounter{appendixc}
        \setcounter{figure}{0}
        \setcounter{table}{0}
        \setcounter{lemma}{0}
        \setcounter{theorem}{0}
        \setcounter{corollary}{0}
        \setcounter{definition}{0}
        \setcounter{equation}{0}
        \renewcommand{\thefigure}{\Alph{appendixc}.\arabic{figure}}
        \renewcommand{\thetable}{\Alph{appendixc}.\arabic{table}}
        \renewcommand{\theappendixc}{\Alph{appendixc}}
        \renewcommand{\thelemma}{\Alph{appendixc}.\arabic{lemma}}
        \renewcommand{\thetheorem}{\Alph{appendixc}.\arabic{theorem}}
        \renewcommand{\thedefinition}{\Alph{appendixc}.\arabic{definition}}
        \renewcommand{\thecorollary}{\Alph{appendixc}.\arabic{corollary}}
        \renewcommand{\theequation}{\Alph{appendixc}.\arabic{equation}}
        \noindent{\tenbf Appendix \theappendixc #1}\par\vspace{5pt}}
\newcommand{\subappendix}[1] {\vspace{12pt}
        \refstepcounter{subappendixc}
        \noindent{\bf Appendix \thesubappendixc. {\kern1pt \bfit #1}}
	\par\vspace{5pt}}
\newcommand{\subsubappendix}[1] {\vspace{12pt}
        \refstepcounter{subsubappendixc}
        \noindent{\rm Appendix \thesubsubappendixc. {\kern1pt \tenit #1}}
	\par\vspace{5pt}}

\newcommand{\textlineskip}{\baselineskip=13pt}
\newcommand{\smalllineskip}{\baselineskip=10pt}

\def\eightcirc{
\begin{picture}(0,0)
\put(4.4,1.8){\circle{6.5}}
\end{picture}}
\def\eightcopyright{\eightcirc\kern2.7pt\hbox{\eightrm c}}

\newcommand{\copyrightheading}[1]
	{\vspace*{-2.5cm}\smalllineskip{\flushleft
	{\eightrm International Journal of Modern Physics D, #1}\\
	{\eightrm $\eightcopyright$\, World Scientific Publishing
	 Company}\\
	 }}

\def\abstracts#1#2#3{{
	\centering{\begin{minipage}{4.5in}\baselineskip=10pt\eightrm
	\centerline{ABSTRACT}
	\parindent=0pt #1\par
	\parindent=15pt #2\par
	\parindent=15pt #3
	\end{minipage} }\par}}

\newcommand{\bibit}{\nineit}

\renewenvironment{thebibliography}[1]			%ALL CHANGES DD 13/3/92
	{\ninerm
	 \baselineskip=11pt				%changed by cheng
	 \begin{list}{\arabic{enumi}.}
	{\usecounter{enumi}\setlength{\parsep}{0pt}
	 \setlength{\leftmargin 17pt}{\rightmargin 0pt}	%changed by cheng
	 \setlength{\itemsep}{0pt} \settowidth		%changed by cheng
	{\labelwidth}{#1.}\sloppy}}{\end{list}}

\newcounter{itemlistc}
\newcounter{romanlistc}
\newcounter{alphlistc}
\newcounter{arabiclistc}

\newcommand{\fcaption}[1]{
        \refstepcounter{figure}
        \setbox\@tempboxa = \hbox{\eightrm Fig.~\thefigure. #1}
        \ifdim \wd\@tempboxa > 5in
           {\begin{center}
        \parbox{5in}{\eightrm \smalllineskip Fig.~\thefigure. #1 }
            \end{center}}
        \else
             {\begin{center}
             {\eightrm Fig.~\thefigure. #1}
              \end{center}}
        \fi}

\newcommand{\tcaption}[1]{
        \refstepcounter{table}
        \setbox\@tempboxa = \hbox{\eightrm Table~\thetable. #1}
        \ifdim \wd\@tempboxa > 5in
           {\begin{center}
        \parbox{5in}{\eightrm\smalllineskip Table~\thetable. #1 }
            \end{center}}
        \else
             {\begin{center}
             {\eightrm Table~\thetable. #1}
              \end{center}}
        \fi}

\def\@citex[#1]#2{\if@filesw\immediate\write\@auxout	%IJMPA, IJMPB ONLY
	{\string\citation{#2}}\fi			%TO DELETE PERCENTAGE
\def\@citea{}\@cite{\@for\@citeb:=#2\do			%KEY WHEN USING
	{\@citea\def\@citea{,}\@ifundefined		%DD 13/3/92
	{b@\@citeb}{{\bf ?}\@warning
	{Citation `\@citeb' on page \thepage \space undefined}}
	{\csname b@\@citeb\endcsname}}}{#1}}

\newif\if@cghi
\def\cite{\@cghitrue\@ifnextchar [{\@tempswatrue
	\@citex}{\@tempswafalse\@citex[]}}
\def\citelow{\@cghifalse\@ifnextchar [{\@tempswatrue
	\@citex}{\@tempswafalse\@citex[]}}
\def\@cite#1#2{{$\null^{#1}$\if@tempswa\typeout
	{IJCGA warning: optional citation argument
	ignored: `#2'} \fi}}

\def\pmb#1{\setbox0=\hbox{#1}
	\kern-.025em\copy0\kern-\wd0
	\kern.05em\copy0\kern-\wd0
	\kern-.025em\raise.0433em\box0}

\def\fnt#1#2{\footnotetext{\kern-.3em
	{$^{\mbox{\scriptsize #1}}$}{#2}}}

\def\fpage#1{\begingroup
\voffset=.3in
\thispagestyle{empty}\begin{table}[b]\centerline{\footnotesize #1}
	\end{table}\endgroup}

\def\runninghead#1#2{\pagestyle{myheadings}
\markboth{{\eightit{\quad #1}}\hfill}{\hfill{\eightit{#2\quad}}}}

\font\tenbf=cmbx10
\font\tenit=cmti10
\font\tenit=cmti10
\font\bfit=cmbxti10 at 10pt
 1
 1
 1

\font\ninerm=cmr9
\font\nineit=cmti9

\font\eightrm=cmr8
\font\eightit=cmti8

\def\qed{\hbox{${\vcenter{\vbox{                          %HOLLOW SQUARE
   \hrule height 0.4pt\hbox{\vrule width 0.4pt height 6pt
   \kern5pt\vrule width 0.4pt}\hrule height 0.4pt}}}$}}

\runninghead{Cosmological Solutions in 2D Poincar\'e Gravity
$\ldots$} {Cosmological Solutions in 2D Poincar\'e Gravity
$\ldots$}

\begin{document}
\normalsize\textlineskip
{\thispagestyle{empty}
\setcounter{page}{1}

\renewcommand{\thefootnote}{\fnsymbol{footnote}} %use symbolic footnote

%print out the publisher copyright heading
\copyrightheading{Vol. 0, No. 0 (1993) 000--000}

\vspace*{0.88truein}

\fpage{1}
\centerline{\bf COSMOLOGICAL SOLUTIONS IN 2D POINCAR\'E GRAVITY}
\vspace{0.37truein}
\centerline{\footnotesize S.N.SOLODUKHIN}
\vspace*{0.015truein}
\centerline{\footnotesize\it
Laboratory of Theoretical Physics, Joint Institute For Nuclear Research}
\baselineskip=10pt
\centerline{\footnotesize\it Head Post Office, P.O.Box 79, Moscow, Russia}
\vspace*{0.21truein}
\abstracts{\noindent
   The 2D model of gravity with zweibeins $e^{a}$ and the Lorentz
connection one-form $\omega^{a}_{\ b}$ as independent gravitational variables
is considered. The solutions of classical equations of motion
which can be interpreted as cosmological ones
are studied.}{}{}
\vspace*{-3pt}\textlineskip
\noindent
\vspace*{0.21truein}

   Numerous recent attempts to formulate the
theory of gravity in the framework of a consistent gauge approach resulted in
constructing the gauge gravity models for the de Sitter and Poincar\'e groups.
 The independent variables are  vielbeins
$e^{a}= e^{a}_{\mu}dx^{\mu}$ and Lorentz connection one-form
$\omega^{a}_{\ b} = \omega^{a}_{\ b,\mu} dx^{\mu}$.
The application of these methods in two dimensions was
justified by attempts to give an alternative description of two-dimensional
dynamical gravity in terms of variables $( e^{a} , \omega^{a}_{\ b} )$. It
was argued also that investigation of simple two-dimensional model leads to
a better understanding of four-dimensional gravity and its
quantization\cite{1,2,3}.
The classical equations of motion were analyzed in conformal gauge in
Ref.\cite{1} and in light cone gauge
in Ref.\cite{2} and
their exact integrability was demonstrated.
The canonical quantization of the model
was recently studied  in Ref.\cite{3}. In Ref.\cite{4} was shown that
classical equations of motion are exactly integrated in coordinate system
determined
by components of 2D torsion. Here we  study in more details the solutions found
in
Ref.\cite{4} which can be interpreted as the  cosmological ones.

    In two dimensions the gauge gravity is described in terms of zweibeins
$e^{a} = e^{a}_{\mu} dz^{\mu}, a=0,1 $ (the 2D metric on the surface $M^{2}$
has the form $g_{\mu \nu}=e^{a}_{\mu} e^{b}_{\nu} \eta_{ab} $) and Lorentz
connection one-form $\omega^{a}_{\ b} = \omega \varepsilon^{a}_{\ b}, \  \omega
=
\omega_{\mu} dz^{\mu} \  (\varepsilon_{ab} =- \varepsilon_{ba}, \
\varepsilon_{01}=1)$.
The curvature and torsion two-forms
are: $R=d\omega, \ T^{a}=de^{a} + \varepsilon^{a}_{\ b}
\omega \wedge e^{b}$.

   The dynamics of gravitational variables $( e^{a}, \omega)$ is determined by
the action:
\begin{equation}
S= \int\limits_{M^{2}}^{} {\alpha \over 2} \ast T^{a} \wedge T^{a} +
{1 \over 2} \ast R \wedge R - {\lambda^{} \over 4} \varepsilon_{ab} e^{a} e^{b}
\end{equation}
where $\ast$ is the Hodge dualization and $\alpha,\lambda$ are arbitrary
constants
and we fixed constant in front of the second term.

   Let us consider variables $\rho = \ast R$ and $q^{a} = \ast T^{a}$.
Variation of action (1) with respect to zweibeins $e^{a}$ and Lorentz
connection $\omega$ leads to the following equations of motion:
\begin{equation}
d \rho =- {\alpha} q^{a}\varepsilon_{ab}e^{b}
\end{equation}
\begin{equation}
\nabla q^{a} = -{1 \over 2{\alpha} } [\rho^{2} +\alpha q^{2}
- \alpha^{2} - \Lambda \alpha] \varepsilon^{a}_{\ b} e^{b},
\end{equation}
where $\nabla q^{a} \equiv dq^{a} + \omega \varepsilon^{a}_{\ b}q^{b}$;
here $q^{2}=q^{a}q^{b} \eta_{ab} \ (\eta_{ab}=diag(+1,-1))$. In (2), (3) the
following
notation was introduced: $\Lambda= {\lambda \over \alpha}- {\alpha}$.

    One particular solution of (2)-(3) is evident.
Assuming $q^{2}=constant$ one gets from (2)-(3), provided
$e^{a}$ are linearly independent everywhere on $M^{2}$:
\begin{equation}
\rho = \pm \alpha^{}, \ q^{a}=0
\end{equation}
in all points of the two-dimensional manifold. That is, torsion is zero
and $M^2$ is the de Sitter space.

In Ref.\cite{4} was shown that if $q^{2}$ be non zero identically everywhere
in $M^{2}$ then the general solution of (2), (3) is easily found in coordinate
system determined by components of 2D torsion. More exactly, assuming
(for definiteness) that $q^{2}=(q^{0})^{2}- (q^{1})^{2} >0$, one can write the
torsion components in the form: $q^{0}=q \cosh \phi, \ q^{1}=q \sinh \phi$.
It follows from (2), (3) that $q^{2}$ is function of curvature $\rho$:
\begin{equation}
q^{2}(\rho) = - {1 \over \alpha} (\rho + \alpha)^{2} + \Lambda
+ \epsilon e^{\rho \over \alpha^{}},
\end{equation}
where $\epsilon $ is integrating constant, which is proportional
to ADM mass.
So we may consider $\rho, \phi$ as the new local coordinates on $M^{2}$. Then
as
shown in Ref.\cite{4} the metric has the form:
\begin{equation}
ds^{2}=
q^{2}(\rho) \exp{(-\frac{2\rho}{\alpha})}(d\phi)^{2}
-\frac{1}{\alpha^{2} q^{2}(\rho)}(d\rho)^{2}
\end{equation}
where $q^{2}(\rho)$ is known function (5). This metric can be rewritten in
Schwarzschild like form.
Indeed, let change variable: $r=e^{-{\rho \over \alpha}}$. Then the
metric (6) takes the form:
\begin{eqnarray}
&&ds^{2}= g(r)dt^{2} - {1 \over g(r)} dr^{2} \nonumber \\
&&g(r)=\epsilon r + \Lambda r^{2} - \alpha r^{2} (1-\ln r)^{2} .
\end{eqnarray}
Generally (7) describes the black hole of charged type with mass $M=
{\epsilon \alpha \over 2}$ embedded in the de Sitter space-time\cite{4}.
If $g(r) \leq 0$ everywhere then metric (7) is the nonstationary but
homogeneous one
($\rho$ is the time coordinate in this case) and describes 2D
cosmology. One may identify in this case points $\phi=0$ and $\phi=\phi_{0}$.
The metric (7) describes then the space-time of topology
$S^{1} \oplus R$. The solutions with such a topology are standard subject for
canonical quantization\cite{3}. Therefore it is of interest to see
what they describe on the classical level.
Since $r$ is time coordinate we may denote: $r \rightarrow t$. Then
metric (7) can be written as follows:
\begin{equation}
ds^{2}={1 \over f(t)} dt^{2}-f(t) d\phi^{2}
\end{equation}
where $f(t)=-g(t)$. This metric can be written in standard "cosmological" form:
\begin{equation}
ds^{2}= d\tau^{2}- a(\tau) d\phi^{2},
\end{equation}
where the cosmological time $\tau$ and cosmic scale factor $a(\tau)$ are
defined as follows:
\begin{equation}
\tau=\int\limits_{}^{t}{ dt' \over \sqrt{f(t')}} \ ; \ a(\tau)=f(t(\tau))
\end{equation}
The metric (9) describes the  expanding of closed  one-dimensional Universe.
The eq.(5) and equation $\rho=-\alpha \ln t(\tau)$ show the evolution of
torsion
and curvature with cosmological time $\tau$.

We begin with consideration of 2D de Sitter solution (4).
Let us choose coordinates $(t, \phi)$ in which the metric takes the form (8).
Then condition (4) leads to equation: $f''(t)= \pm 2\alpha$. We will consider
here
the case $\rho=-\alpha >0$. Then $f(t)$ takes the form:
\begin{equation}
f(t)=\alpha(t-t_{1})(t-t_{2})
\end{equation}
where $t_{1} \leq t_{2}$. The coordinate $\tau$ can be determined only in the
region $t \geq t_{2}$. From (10) we obtain:
\begin{eqnarray}
&&\tau={1 \over \sqrt{\alpha}} \ln (t- {(t_{1}+t_{2}) \over 2} +
\sqrt{(t-t_{1})(t-t_{2})} ), \nonumber \\
&&a(\tau)=\alpha (t(\tau) - t_{1})(t(\tau) -t_{2})
\end{eqnarray}
We see that if $t_{1} \neq t_{2}$ then the Universe starts the expanding from
zero
radius ($a=0$) at $\tau= {1 \over \sqrt{\alpha}} \ln ({{t_{2}-t_{1}} \over 2})$
and expands exponentially such that for large $\tau$: $a(\tau) \approx
{\alpha \over 4} exp( 2 \sqrt{\alpha} \tau)$. The  eqs.(12) takes simplest
form if $t_{2}=t_{1}$:
\begin{equation}
\tau={1 \over \sqrt{\alpha}} \ln (2(t -t_{1})) \ ; \ a(\tau)= {\alpha
\over 4} exp( 2 \sqrt{\alpha} \tau)
\end{equation}
The Universe expands now exponentially from infinite past ($\tau =- \infty, \
a=0$) to
infinite future ($\tau= + \infty$). Thus the constant curvature solution (4)
describes
the ordinary inflation.

Let us consider now the general solution with nontrivial torsion (7). We will
restrict ourselves to the case $\Lambda=0$. Let now $\epsilon=0, \alpha >0$.
Then (7)   takes the form (8) with $f(t)=\alpha t^{2}(1- \ln t)^{2}$.
The view of functions $\tau (t) , \ a(\tau)$ depends in what interval
variable $t$
changes. Let $t \geq e$ ($e$ is Euler number) then we obtain:
\begin{equation}
\tau= {1 \over \sqrt{\alpha}} \ln (\ln t -1) \ ; \
a(\tau)=\alpha \exp [2 (1+e^{\sqrt{\alpha} \tau}+ \sqrt{\alpha} \tau)]
\end{equation}
The Universe starts evolution from constant curvature $(\rho=-\alpha)$
configuration of zero radius $(a=0)$ at infinite past ($\tau=- \infty)$.
Then it expands with law $exp(exp \sqrt{\alpha} \tau)$ and such a stage can be
called
"super-inflation".

If $0 < t \leq e$ then we get:
\begin{equation}
\tau=- {1 \over \sqrt{\alpha}} \ln (1- \ln t) \ ; \
a(\tau)=\alpha \exp [2 (1-e^{- \sqrt{\alpha} \tau}- \sqrt{\alpha} \tau)]
\end{equation}
The metric (9)  with (15) describes then  expanding of the Universe from
singular
state ($\rho= + \infty$) of zero radius in infinite past $( \tau =- \infty)$.
It has maximal
radius ($a=\alpha$) at $\tau =0$ and then collapses to constant curvature
($\rho=-\alpha$) configuration with $a=0$ in infinite future $(\tau =+
\infty)$.

Let now $\epsilon <0$ then
\begin{equation}
f(t)=|\epsilon| t + \alpha t^{2}(1- \ln t)^{2}
\end{equation}
For small $t \stackrel{\sim}{>}  0$ we obtain from (10) that evolution of the
Universe is determined by
$\epsilon$-term: $\tau \approx 2 \sqrt{{t \over |\epsilon |}} ; \ a \approx
{ \epsilon^{2} \over 4} \tau^{2}$,  while for large $t$ (and consequently
for large $\tau$) the second term in (16) dominates and we obtain the
super-inflationary
stage as before (14). The corresponding evolution of curvature is following:
$\rho=-\alpha \ln ({\epsilon \over 4} \tau^{2})$ for small $\tau$ and
$\rho=- \alpha exp( \sqrt{\alpha} \tau)$ for large cosmic time $\tau$.
The $\epsilon$-term in (7), (16) is due to the presence of non-gravitating
matter with energy-momentum
tensor $T_{00} \sim \epsilon \delta (\tau)$ in the initial state of the
Universe. So
we may interpret this result as that the matter determines the evolution of the
Universe
at the beginning of  the expanding while for later cosmological times the
self-gravitating
forces dominate.

Concluding we considered the solutions of cosmological type in 2D Poincare
gravity.
The different possible regimes are analyzed. The most interesting one
 is stage of super-inflation when the scale factor $a(\tau)$ changes as
$exp (exp \sqrt{\alpha} \tau)$ with cosmological time $\tau$.

\nonumsection{References}


\noindent
\begin{thebibliography}{000}
\bibitem{1} M.O.Katanaev, I.V.Volovich, {\bibit Ann.Phys.} {\bf 197} (1990), 1;
            M.O.Katanaev, {\bibit J.Math.Phys.} {\bf 31} (1991), 2483;
            {\bf 34} (1993), 22.
\bibitem{2} W.Kummer, D.J.Schwarz, {\bibit Phys.Rev.} {\bf D45} (1992), 3628.
\bibitem{3} P.Schaller, T.Strobl, {\bibit Canonical quantization of
non-Einsteinian
            gravity and the problem of time}, TUW-92-13.

\bibitem{4} S.Solodukhin, {\bibit JETP Lett.}, {\bf 57} (1993), 329; {\bibit
Phys.Lett.}
            {\bf B}, (to appear).
\end{thebibliography}
\end{document}